\documentclass[conference,10pt, a4paper]{IEEEtran}
\usepackage{xcolor}
\usepackage{mathtools}
\usepackage{epsfig,makeidx,color}
\usepackage{amsmath,amssymb,bbm,enumitem}
\usepackage{cite,graphicx,lipsum}
\usepackage[switch,pagewise]{lineno}
\usepackage{hyperref}
\usepackage[caption=false,font=footnotesize]{subfig}
\usepackage{algorithm}
\usepackage{algorithmic}
\usepackage{acronym}
\usepackage{array}    
\hypersetup{
        colorlinks = true,
        citecolor=red,
}
\usepackage{tikz}

\def\cH{{\cal H}}

\def\cG{{\cal G}}

\def\be{ \begin{equation} }
\def\ee{ \end{equation} }
\def\bea{ \begin{eqnarray} }
\def\eea{ \end{eqnarray} }

\def\b0{{\bf 0}}

\def\cC{{\cal C}}

\def\cD{{\cal D}}
\def\cF{{\cal F}}
\def\cB{{\cal B}}

\def\cI{{\cal I}}

\def\cW{{\cal W}}

\acrodef{TX}{transmitter}
\acrodef{RX}{receiver}
\acrodef{BEC}{bit erasure channel}
\acrodef{MSE}{mean squared error}
\acrodef{LPIPS}{learned perceptual image patch similarity}
\acrodef{AIGC}{AI-generated content}
\acrodef{ML}{machine learning}
\acrodef{NLP}{natural language processing}
\acrodef{LLM}{large language model}
\acrodef{CLIP}{contrastive language-image pre-training}
\acrodef{SemCom}{semantic communication}
\acrodef{SemPA}{semantic packet aggregation}
\acrodef{SemRT}{semantic repeated transmission}
\acrodef{GA}{Genetic Algorithm}
\acrodef{OOV}{out-of-vocabulary}
\acrodef{PA}{packet aggregation}
\acrodef{ATS}{average token similarity}
\acrodef{SemPA-GBeam}{SemPA with genetic beam search}

\ifCLASSOPTIONonecolumn
\else

\fi

\usepackage{geometry}
\graphicspath{ {./images/} }
\usepackage{graphicx}
\begin{document}


\title{Semantic Packet Aggregation for Token Communication via Genetic Beam Search}
\author{Seunghun Lee, Jihong Park, Jinho Choi, and Hyuncheol Park}

\maketitle

\begin{abstract} 
Token communication (TC) is poised to play a pivotal role in emerging language-driven applications such as AI-generated content (AIGC) and wireless language models (LLMs). However, token loss caused by channel noise can severely degrade task performance. To address this, in this article, we focus on the problem of semantics-aware packetization and develop a novel algorithm, termed semantic packet aggregation with genetic beam search (SemPA-GBeam), which aims to maximize the average token similarity (ATS) over erasure channels. Inspired from the genetic algorithm (GA) and the beam search algorithm, SemPA-GBeam iteratively optimizes token grouping for packetization within a fixed number of groups (i.e., fixed beam width in beam search) while randomly swapping a fraction of tokens (i.e., mutation in GA). Experiments on the MS-COCO dataset demonstrate that SemPA-GBeam achieves ATS and LPIPS scores comparable to exhaustive search while reducing complexity by more than 20×.
\end{abstract}


\ifCLASSOPTIONonecolumn
\baselineskip 28pt
\fi
\section{Introduction}


Recent advances in large language models (LLMs) and generative AI (GenAI) have enabled emerging applications such as \ac{AIGC} \cite{Minrui24} and wireless LLMs~\cite{oh2025HLM}, which rely on communicating text-based tokens, referred to as token communication (TC) \cite{qiao2025tc}. These tokens convey the semantics of raw data (e.g., generated images in AIGC), and therefore token loss due to channel noise can significantly impair the effectiveness of their target task. This motivates us to revisit packetization strategies for TC, with an emphasis on token semantics.

In classical communication systems, where all bits are treated equally, optimal packetization boils down to a packet length optimization problem. In contrast, accounting for the semantics of tokens within packets, we view optimal packetization as a token grouping problem. In our prior work \cite{lee2025sempa}, we addressed this by maximizing \ac{ATS}, defined as the average cosine similarity between the original message and the received packets over erasure channels. This ATS maximization problem is NP-hard, and our previous solution in \cite{lee2025sempa} employed a greedy combinatorial optimization algorithm, termed \ac{SemPA}, which incurred high computational complexity.

In this paper, we propose a novel algorithm for ATS maximization in TC, inspired by \ac{GA} and beam search, coined \ac{SemPA-GBeam}. Specifically, given a fixed number of packets and packet length, \ac{SemPA-GBeam} begins with a limited set of randomly initialized token groupings (analogous to a fixed beam width in beam search) and sequentially selects the best combination over a fixed number of iterations (similar to a fixed depth). Although ATS maximization is inherently non-sequential, we enable this iterative refinement by introducing random token swaps between packets (akin to mutation in GA). Using the MS-COCO image-caption dataset \cite{MSCOCO14}, our simulations demonstrate that \ac{SemPA-GBeam} achieves ATS and Learned Perceptual Image Patch Similarity (LPIPS) scores that closely match those of exhaustive search, under both word-level and subword-level tokenization, while reducing computational complexity by more than 20 times compared to full search.

Note that SemPA is related to recent studies on text-based semantic communication \cite{Qin22,Xidong23,Jinsong24} as well as LLM-based communication techniques for source and channel coding \cite{Nam2023LanguageOrientedCW}, error correction \cite{Guo22,Guo24}, and multimodal data transmission \cite{cicchetti24}. However, these works do not address the semantics-aware packetization problem that \ac{SemPA} focuses on. 
It is also worth noting that semantic-importance based text communication and the impact of single-word loss on semantic degradation have been studied in \cite{Guo22,Guo24}. Nonetheless, this line of work does not analyze multi-word or multi-token semantic loss, which is the primary focus of SemPA.

\begin{figure*}[t]
\centering
\includegraphics[width=0.9\textwidth]{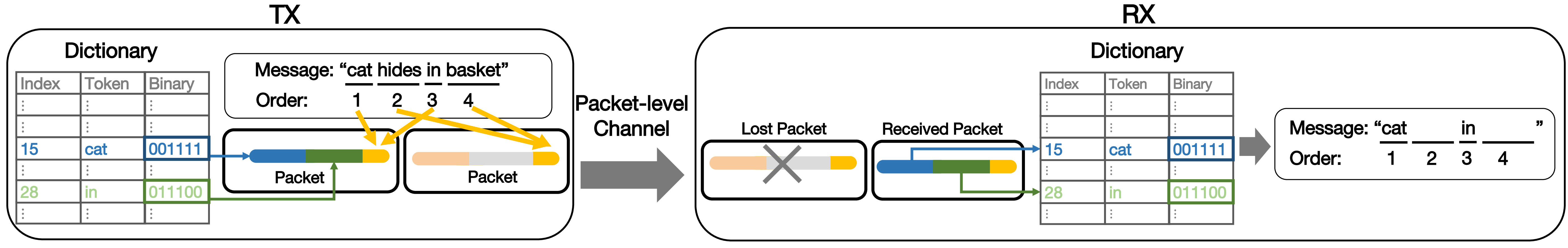}
\vspace{-0.25cm}
\caption{The structure of token communication system. When tokens are packetized based on their semantics, the received packets contain important semantics.}\label{fig:1}
\end{figure*}

\section{System Model}
\subsection{Token Communication Scenario}
Fig.~\ref{fig:1} illustrates a semantic communication scenario where a text sentence consisting of \(K\) tokens, \([w_1, w_2, \cdots, w_K]\), is encoded and transmitted from a \ac{TX} to a \ac{RX} over an unreliable packet-level channel. Here, $w_k$ denotes the $k$th token. In practice, these tokens may be derived using word-based tokenization, where each token corresponds to a word delimited by whitespace, or via a subword-based tokenization method (e.g., BPE, WordPiece) that partitions the text into smaller, learned units.

In this scenario, a high compression ratio renders the reconstruction at the \ac{RX} highly sensitive to channel errors; that is, if some tokens are lost, the \ac{RX} might fail to reconstruct a text sentence that carries the same meaning as the original. To mitigate the risk of losing important semantics due to packet loss, the text sentence is partitioned into multiple packets such that essential semantic information is maintained even if some packets are lost, where each packet contains \(M\) tokens.
We denote the packets by \(\cC_1, \cC_2, \cdots, \cC_{N}\), where \(N = \frac{K}{M}\) is derived from the total number of tokens \(K\) in the sentence and the packet length \(M\). We define the set \(\cG = \{\cC_1, \cC_2, \cdots, \cC_{N}\}\) as the packet group that contains all \(N\) packets necessary to convey the complete meaning of the sentence. 
We assume that the original ordering of tokens is fully recoverable at the \ac{RX} through the packet header information.
In Section~\ref{sec:GBeam_ATS_problem}, we present methods to form the packet group to be robust against channel errors.

We consider a packet-level channel model, since network layer channels are typically modeled as packet-level channels \cite{Sundararajan11}.
In the channel, we assume that a packet is lost with probability \(p\). Let \({\hat{\cC}}\) represent the packet received at the \ac{RX}, which is defined as 
\begin{align}\label{Z}
{\hat{\cC}} = \begin{cases}
{\cC}, & \text{with probability } 1 - p, \\
\emptyset, & \text{with probability } p.
\end{cases}
\end{align}
In \eqref{Z}, we define packet loss as receiving the empty set.


\subsection{Expected Similarity Score between Different Messages}
The goal is to form \(\cC \in \cG\) and transmit each of them as a packet sequentially, so that the received packets contain the important semantic of the message. 
To this end, firstly, we introduce a metric that measures the similarity between semantics in texts based on pre-trained model \cite{CLIP}. In particular, the following cosine similarity score, which is widely used to measure text similarity \cite{Qin22}, \cite{Guo22}, is considered:
\begin{align}\label{sim_f}
    \phi(x, y) = \frac{g(x)^{\mathrm{T}} g(y)}{||g(x)|| \cdot ||g(y)||},
\end{align}
where \(\phi(\cdot, \cdot)\) is the normalized cosine similarity function which needs two arguments as input and \(g(\cdot)\) denotes the text encoding process of the pre-trained model, transforming inputs into feature vectors.


From the channel model and text similarity measurement metric, we can introduce the \ac{ATS}, which averages the similarities between the received message delivered through the unreliable channel and the original transmitted message. Letting \(\cF(\cG) =  \cup_{\cC \in \cG}\cC\), the \ac{ATS} is defined as
\begin{align}
\hspace{-0.5pc}{\mathbb E}\left[\phi(\cF(\cG), \cW)\right] = \sum_{\cH \subseteq \cG}\;(1 - p)^{|\cH|}p^{N - |\cH|}\cdot \phi\bigl(\mathcal{F}(\mathcal{H}), \cW\bigr),
\end{align}
where \(\cH\) represents any possible subset of \({\cG}\).
\section{Genetic Beam Search based Semantic Packet Aggregation}\label{sec:GBeam_ATS_problem}
\subsection{Problem Formulation}\label{sec:ATS_problem}
In this section, we formulate the problem of maximizing the \ac{ATS} in the presence of packet losses during the transmission of $\cG$. 
The \ac{ATS} maximization problem can be formulated as:
\refstepcounter{equation}\label{P0}
\begin{flalign*}
\textbf{P1: }&\max \limits_{\cG}{\mathbb E}\left[\phi(\cF(\cG),\cW)\right]\tag{\theequation a}\label{P0a} &&
\\& {~{\text {s.t.}}}\hspace{0.3pc} \cF({\cG})\!=\!{\cW}\;({\text{complete transmission constraint}}),\hspace{-0.5pc}\tag{\theequation b}\label{P0b}&&
\\& \hphantom{~{\textrm {s.t.}}}\hspace{0.3pc}|\cC| = M\hspace{0.3pc}\forall \cC \in \cG \;({\text{packet length constraint}}). \tag{\theequation c}\label{P0c}&&
\end{flalign*}
The length of the packets is constrained as \eqref{P0c}.

Finding the optimal \(\cG\) in {\textbf{P1}} is computationally challenging due to the presence of packet dependencies. 
The final message reconstruction relies on the aggregated information from the received packets.
As a result, the optimal choice of tokens in each packet depends on how previous packets were composed and transmitted. Due to such dependency, to convey the semantics of the message through the packets, in \ac{TX}, $\cC \in \cG$ should be optimized collectively.

A core obstacle in optimizing $\cG$ instead of $\cC \in \cG$ is the exponential number of possible packet groups. Since the objective function involves the non-linear metric $\phi(\cdot, \cdot)$ derived from a language model $g(\cdot)$, there is no straightforward closed-form or polynomial-time approach for optimizing $\cG$ in \eqref{P0a}.
If we attempt to apply a full search (i.e., Brute-Force Search), every possible partition of \(K\) tokens into packets of size \(M\) must be enumerated, resulting in a total of \(\frac{K!}{(M!)^{N}\cdot N!}\) groups. This number grows intractably as \(K\) and \(N\) increase.

In fact, the dominant computational burden arises from the text encoding steps required to evaluate \(g(\cdot)\) for computing \(\phi(\cdot,\cdot)\). While one could precompute and store the similarity values for all \(2^K\) subsets \(\cH \subset \cW\) in a dictionary—thereby avoiding repeated text encoding during the computation of the ATS for each feasible \(\cG\)—the construction of such a dictionary itself necessitates \(2^K\) text encoding steps, which is prohibitively high. Consequently, we define computational complexity as the number of text encoding steps required and focus on reducing the number of text encoding steps, which serves as the primary measure of computational complexity in our analysis.
In this context, we introduce \ac{SemPA-GBeam} method in Section~\ref{sec:GBeam} that can find a near-optimal solution for \eqref{P0} with reduced computational complexity compared to full search.

\subsection{Genetic Beam Search}\label{sec:GBeam}
To mitigate the computational complexity inherent in a full search approach, a standard \ac{GA} method can be adopted. This well-established evolutionary algorithm iteratively improves candidate solutions through the processes of selection, crossover, and mutation, ultimately converging on a near-optimal solution \cite{katoch21}.

To be specific, each chromosome is represented as a random set of $N$ packets of fixed size $M$ drawn from the overall token set $\mathcal{W}$. For each generation, we evaluate the fitness of each chromosome by computing its exact \ac{ATS}.
However, applying standard \ac{GA} requires defining appropriate crossover and mutation operations for packet group representations. A naive crossover that simply swaps packets between two parent chromosomes might fail to satisfy the complete transmission constraint in \eqref{P0b} (e.g., duplicated tokens may arise, resulting in incomplete packet coverage).

An alternative approach inspired by the well-known beam search algorithm provides a promising direction \cite{snell24}. Beam search traditionally operates by selecting a fixed number of promising candidate solutions (beams) at each iteration and expanding each candidate by sampling multiple proposals, thus systematically refining towards near-optimal solutions without relying on crossover. Instead, it generates diversity by branching from a limited number of highly ranked candidates, ensuring solution quality.

However, our \ac{ATS} maximization problem is inherently non-sequential, making direct application of beam search unsuitable, particularly due to the step-wise sampling operations. To effectively adapt this concept to our non-sequential packet grouping scenario, drawing inspiration from the mutation and generation processes in \ac{GA}, we propose \ac{SemPA-GBeam} that employs a mutation-based token swapping strategy, while maintaining a set of candidate packet groups (beams) across iterations, analogous to generations in \ac{GA}.

\begin{figure}[t]
\centering
\includegraphics[width=0.8\columnwidth]{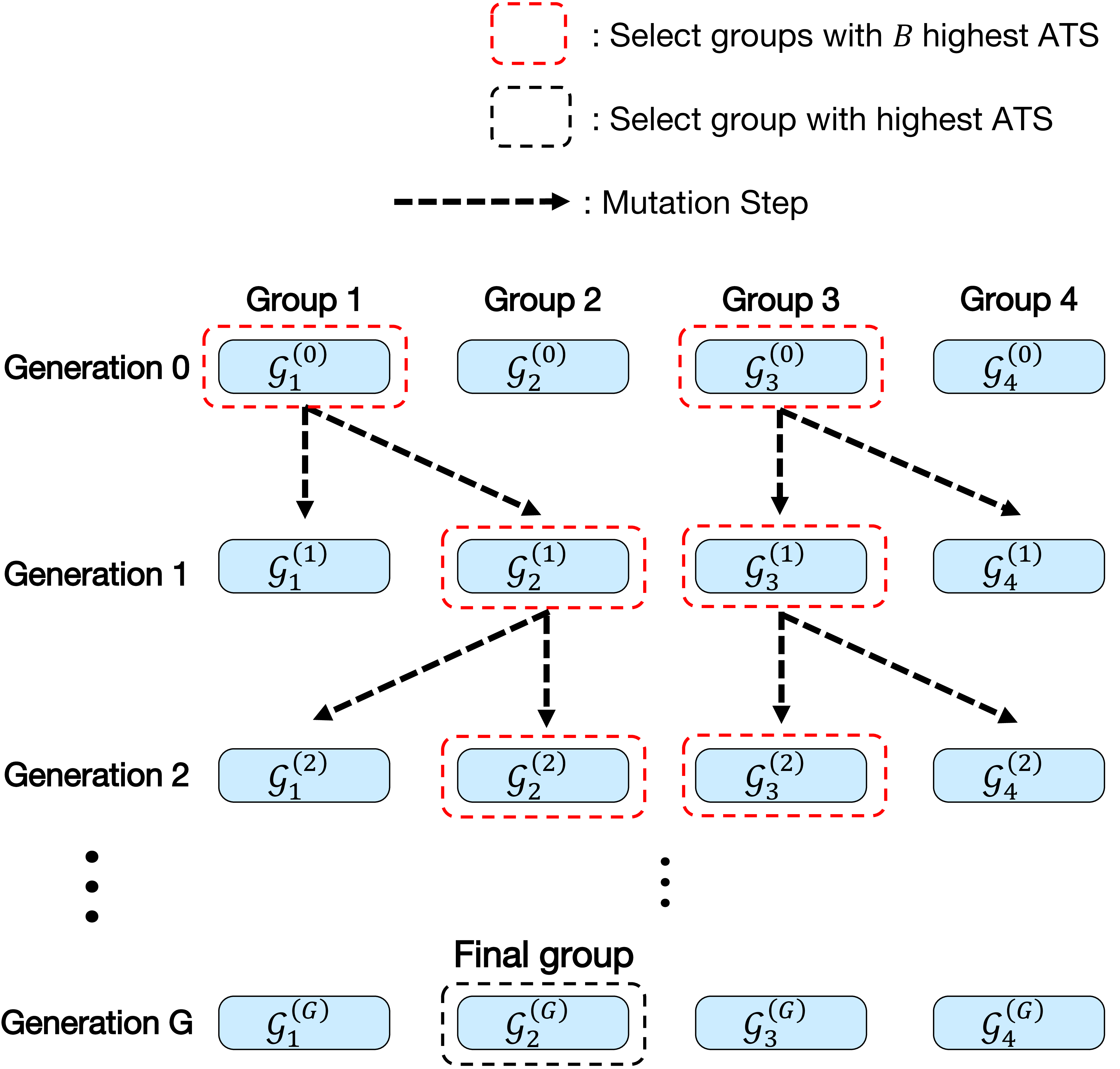}

\caption{An illustration of \ac{SemPA-GBeam}.}\label{fig:GBeam}
\end{figure}

To be specific, the genetic beam search algorithm operates through the following steps:

\begin{enumerate}
\item \textbf{Initialization}.\quad
Randomly generate an initial set of $L$ groups, each comprising $N$ packets of fixed size $M$:
\begin{equation}
\cG_b^{(0)} = \{\cC_{b,1}^{(0)}, \dots, \cC_{b,N}^{(0)}\}, \quad \ell=1,\dots,L.
\end{equation}

\item \textbf{Beam Selection}.\quad
At generation $g$, evaluate the \ac{ATS} of each group and select the top $B$ groups:
\begin{equation}
\cB^{(g-1)} = \text{Top-}B\bigl(\{{\cG_\ell^{(g-1)}}\}_{\ell=1}^L\bigr).
\end{equation}

\item \textbf{Beam Expansion via Mutation}.\quad
From each selected beam ${\cG_{\ell}^{(g-1)}} \in \cB^{(g-1)}$, create $\frac{L}{B}$ candidate groups through token swapping mutations:
\begin{enumerate}
\item \emph{Packet Selection}: Randomly choose two distinct packets $\cC_m$ and $\cC_n$ within the selected beam:
\begin{equation}
\cC_m, \cC_n \subset {\cG_{\ell}^{(g-1)}}, \quad m \neq n.
\end{equation}
\item \emph{Token Swapping}: Randomly select one token from each packet and swap them to form new packets:
\begin{align}
    \cC_m' &= (\cC_m \setminus \{w_{\text{old}}\}) \cup \{w_{\text{new}}\},\\
    \cC_n' &= (\cC_n \setminus \{w_{\text{new}}\}) \cup \{w_{\text{old}}\}.
\end{align}

\item \emph{New Beam Formation}: Replace the original packets with mutated packets to form new candidate beams:
\begin{equation}
    \hat{\cG} = \left\{\cC_1, \dots, \cC_m', \dots, \cC_n', \dots, \cC_N\right\}.
\end{equation}

\end{enumerate}

\item \textbf{Candidate Evaluation and Selection}.\quad
Evaluate the \ac{ATS} of each newly mutated beam candidate. From these candidates, select the top $B$ beams to proceed to the next generation:
\begin{align}
\cB^{(g)} = \text{Top-}B\bigl( {\{\cG_\ell^{(g)}}\}_{\ell=1}^L \bigr).
\end{align}

\end{enumerate}
SemPA-GBeam iterates 1)-4) over $G$ generations using a population size of $L$.

As a result, the total computational complexity of SemPA-GBeam is $G\cdot L\cdot 2^{N}$, which is significantly lower than $2^K$ of full search with moderate $G$ and $L$, as we will demonstrate by simulation in Section~\ref{sec:simulation_results}.
Moreover, genetic beam search integrates the beneficial exploration properties of beam search and mutation strategies from \ac{GA}, effectively addressing the packet grouping problem with substantially reduced computational complexity.

\section{Simulation Results}\label{sec:simulation_results}
The evaluation of the proposed \ac{SemPA-GBeam} is conducted by comparing with baseline methods, using the MS-COCO image-caption dataset \cite{MSCOCO14}. We assume that the tokens of the text sentence \([w_1, w_2, \cdots, w_K]\), which are sampled from the dataset, are included in the pre-defined dictionary \( \cD = \{v_1, v_2, \cdots, v_{|\cD|}\}\) in both \ac{TX} and \ac{RX}. The tokens in each packet are encoded to indices in the dictionary, which needs \(\lceil \log_2 |\cD| \rceil\) bits for each token. 
The cosine similarity is measured by computing cosine of the angles between two text embeddings which are derived from the pre-trained \ac{CLIP} model. 

We consider three \ac{PA} methods for semantic communication. In Random \ac{PA}, the token set \(\cW\) is randomly partitioned into packets of fixed size \(M\), incurring no computational cost but yielding relatively low \ac{ATS}. full search exhaustively enumerates all feasible packet groups that satisfy the constraints and selects the optimal $\cG$ by maximizing the ATS, though each evaluation requires \(2^N\) text encoding steps. For Without Packetization method, all tokens are aggregated into a single packet, which increases the probability of complete transmission of message, but it is highly vulnerable to losing whole tokens.

\begin{figure}
     \centering
     \subfloat[Comparison of \ac{ATS} across different \ac{PA} methods. Word-based tokenization is applied.]{%
         \includegraphics[width=0.48\columnwidth]{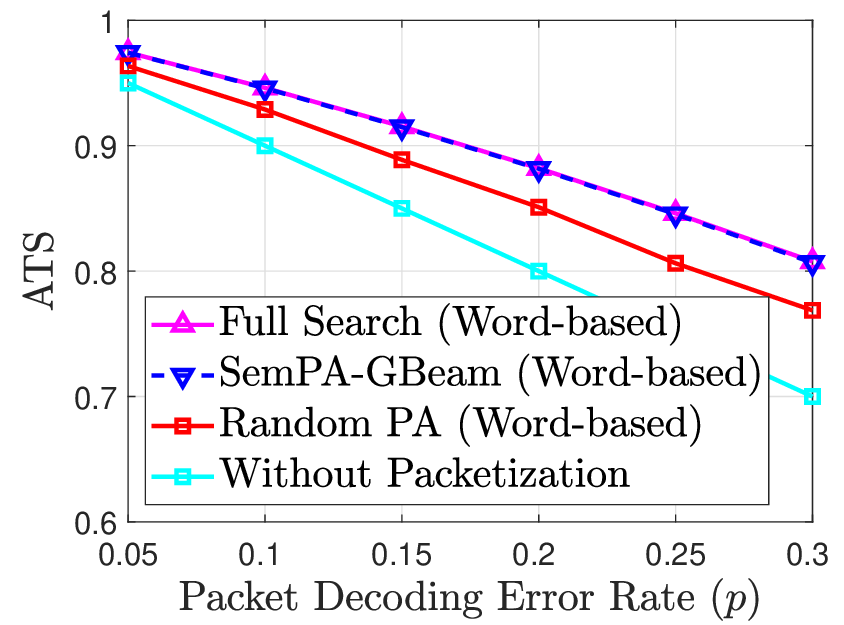}
         \label{fig:plot_p_word}
     }
     \hfill
     \subfloat[Comparison of \ac{ATS} across different \ac{PA} methods. Subword-based tokenization is applied.]{%
         \includegraphics[width=0.48\columnwidth]{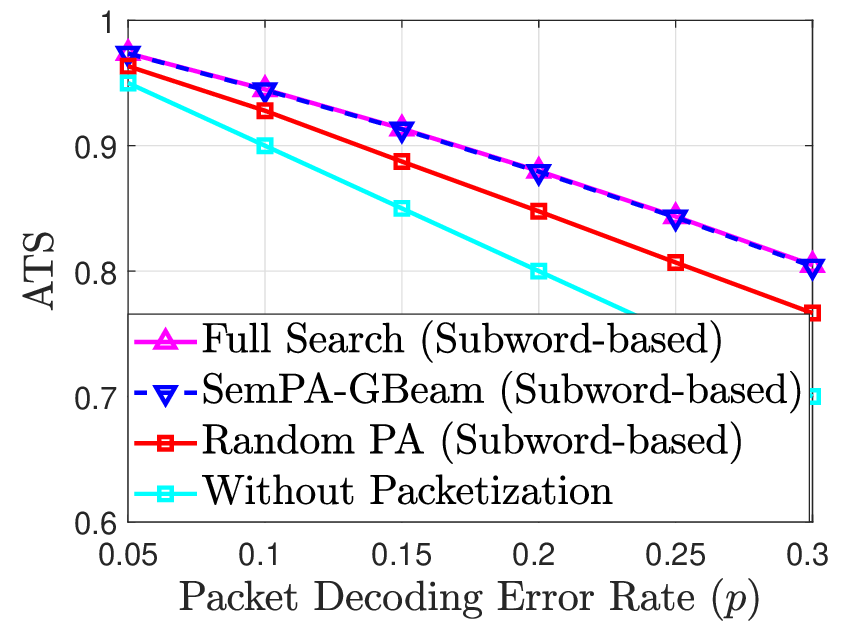}
         \label{fig:plot_p_token}
     }
     \caption{Comparison of \ac{ATS} between original and decoded messages using different packet transmission methods. $K = 8, M = 4, L = 10, G = 5, L = 10, B = 2.$} 
     \label{fig:plot_p}
\end{figure}

As demonstrated in Fig.~\ref{fig:plot_p}, a comparison of \ac{ATS} obtained by different \ac{PA} methods for different $p$ is presented. Fig.~\ref{fig:plot_p_word} shows that the \ac{SemPA-GBeam} method achieves a score of 99.88\%, which is almost equivalent to the optimal full search at $p=0.3$. It is noteworthy that the computational complexity of other methods in Tab.~\ref{tab:complexity} is not a limiting factor for the effectiveness of the \ac{SemPA-GBeam}, which is capable of achieving high similarity scores while being computationally efficient. Furthermore, in Fig.~\ref{fig:plot_p_token}, subword-based tokeniZation is evaluated under the same setup as in Fig.~\ref{fig:plot_p_word}, yielding slightly lower similarity scores than word-based tokenization. In particular, when switching tokenization to subword-based tokenization, the average score of \ac{SemPA-GBeam} experiences a negligible drop to $0.0029$ when $p=0.3$. The reason for this is that subword tokens with higher correlations within the same source word may be scattered into separate packets, and when these packets are erased, the potential loss of semantic representation is more pronounced. Consequently, subword-based tokenization generally results in somewhat lower overall similarity scores.

\begin{figure}
     \centering
     \subfloat[Computational complexities of full search and \ac{SemPA-GBeam} as $K$ increases. $M = 2$, $G = 5$, $L = 10$.]{%
         \includegraphics[width=0.48\columnwidth]{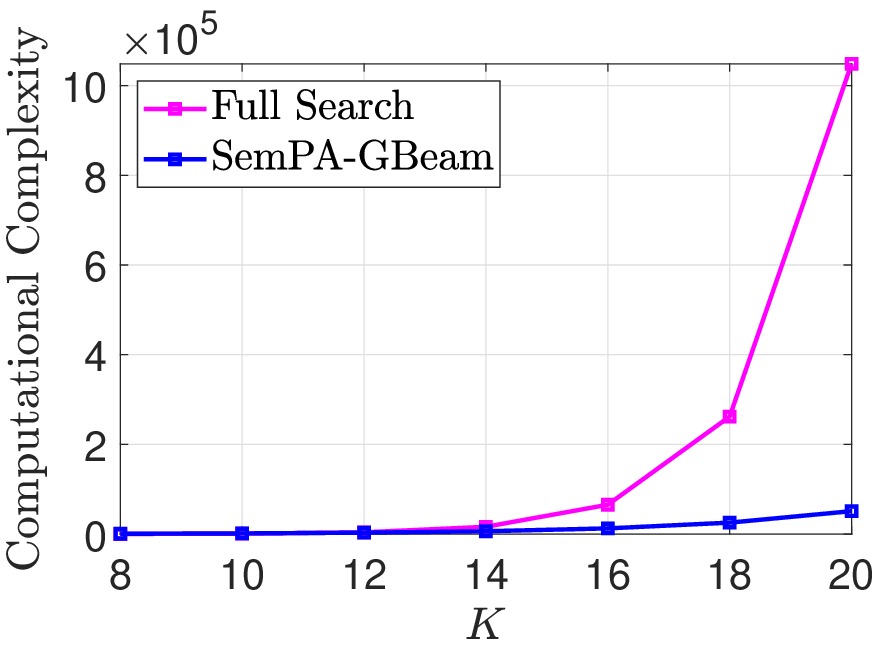}
         \label{fig:Complexity_analysis_K}
     }
     \hfill
     \subfloat[Computational complexities of full search and \ac{SemPA-GBeam} as $M$ increases. $K = 12$, $G = 5$, $L = 10$.]{%
         \includegraphics[width=0.48\columnwidth]{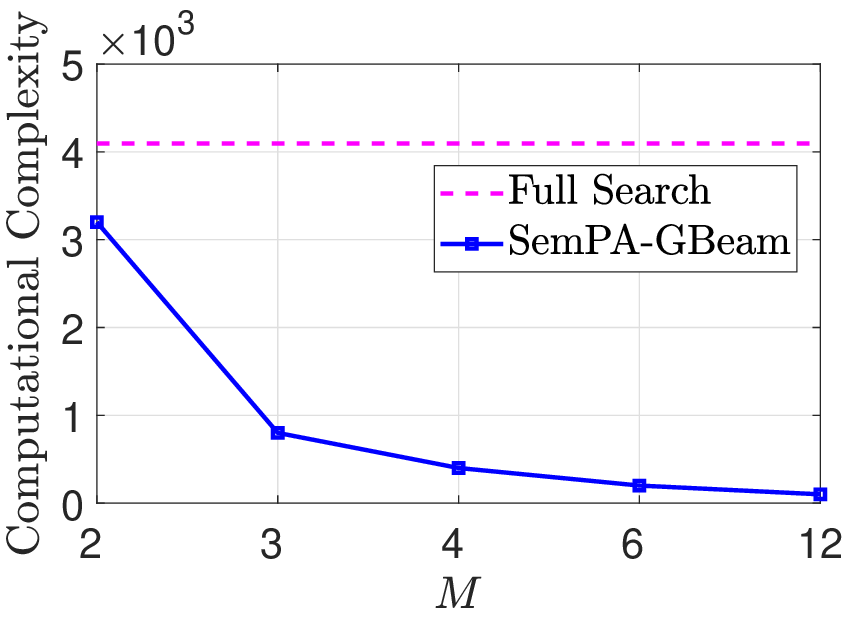}
         \label{fig:Complexity_analysis_M}
     }
     \caption{Comparison of \ac{ATS} between original and decoded messages using different packet transmission methods.} 
     \label{fig:Complexity_analysis}
\end{figure}

Fig.~\ref{fig:Complexity_analysis} compares the computational complexity of different methods. In particular, in Fig.~\ref{fig:Complexity_analysis_K}, at $K=20$, \ac{SemPA-GBeam} requires $20.48$ times lower complexity compared to full search. This is because while the complexity of full search increases exponentially with $K$, the proposed \ac{SemPA-GBeam} scales exponentially with $N = K/M$, as summarized in Tab.~\ref{tab:complexity}. Similarly, Fig.~\ref{fig:Complexity_analysis_M} shows that as $M$ increases, the complexity of \ac{SemPA-GBeam} further decreases, achieving a $20.48$ times reduction over full search at $M=6$. These results demonstrate that by maintaining a moderately large $N$, \ac{SemPA-GBeam} achieves near-optimal performance with significantly lower computational cost.

\begin{table}[t]
\centering
\caption{Computational Complexity for full search vs. SemPA-GBeam}
\label{tab:complexity}
\begin{tabular}{|c|c|}
\hline
\textbf{Method} & \textbf{Complexity} \\
\hline
\hline
full search & $\displaystyle 2^K$ \\
\hline
\ac{SemPA-GBeam} & $\displaystyle G\cdot L\cdot2^{N}$ \\
\hline
\end{tabular}
\end{table}

\begin{figure*}[t]
     \centering
     \subfloat[Comparison of LPIPS across different \ac{PA} methods with word-based tokenization. Example AIGCs generated from captions reconstructed via \ac{SemPA-GBeam} and Random PA.]{%
     \includegraphics[width=0.48\textwidth]{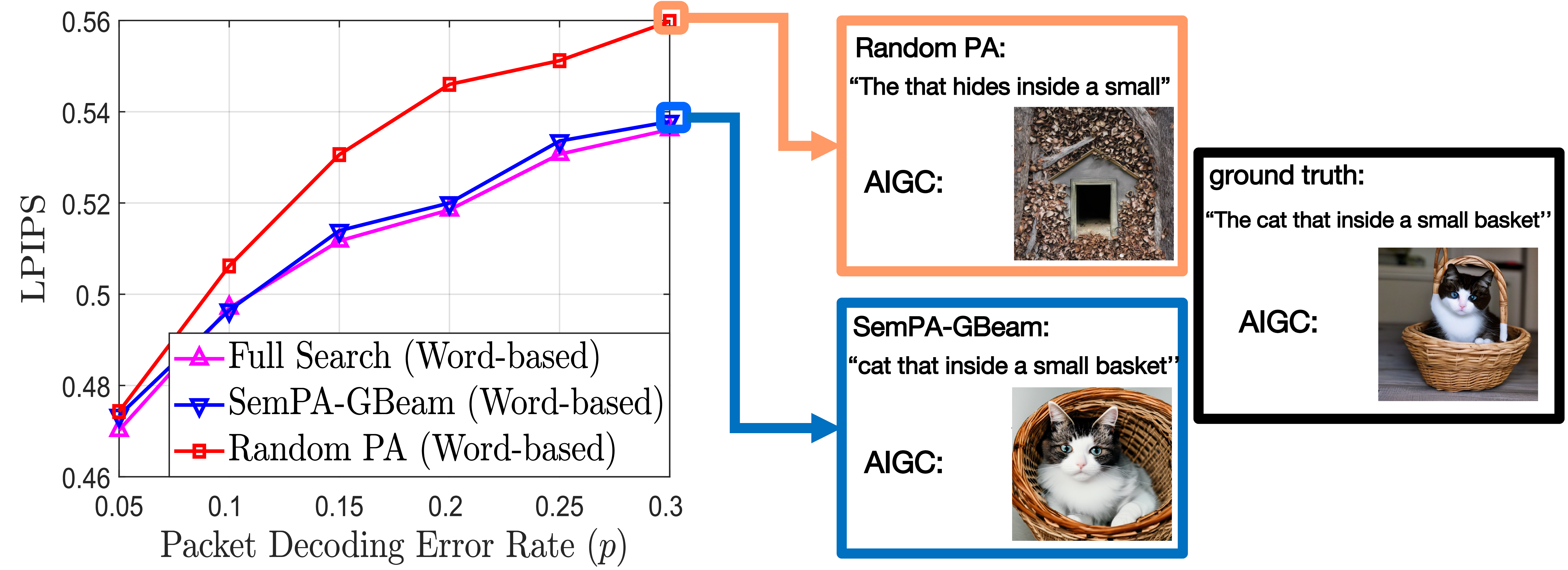}
         \label{fig:plot_word_LPIPS}
     }
     \hfill
     \subfloat[Comparison of LPIPS across different \ac{PA} methods with subword-based tokenization. Example AIGCs generated from captions reconstructed via \ac{SemPA-GBeam} and Random PA.]{%
         \includegraphics[width=0.48\textwidth]{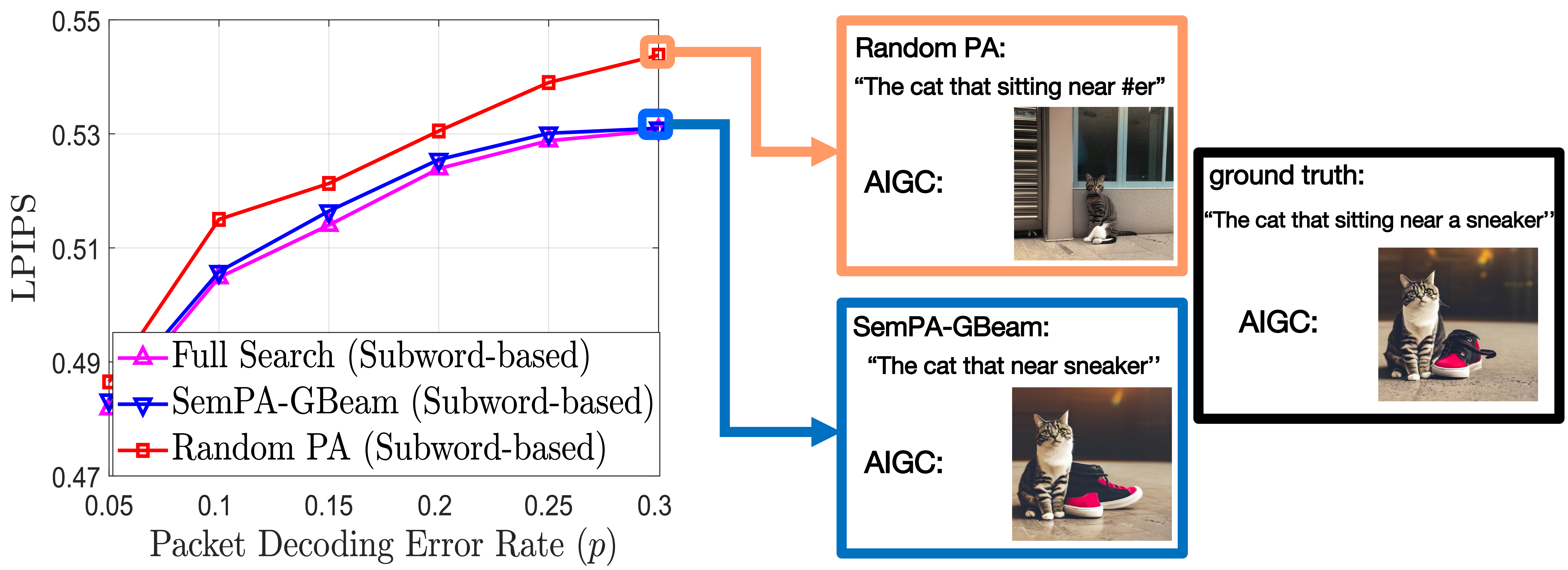}
         \label{fig:plot_token_LPIPS}
     }
     \caption{Average LPIPS between AIGCs from received and original messages across different \ac{PA} methods as $p$ increases. Both tokenization methods show that \ac{SemPA-GBeam} preserves semantic content more effectively than baseline methods, particularly at higher error rates. $K = 8$, $M = 4$, $L = 10$, $G=5$, $B = 2$.} 
     \label{fig:plot_LPIPS}
\end{figure*}

Finally, for image-level performance analysis, we conduct experiments on the MS-COCO image-caption dataset to evaluate the performance of different \ac{PA} methods. In our experimental setup, we first prepare image–caption pairs from the MS-COCO dataset. Then, a coarse version of the original image is generated using canny edge detection, denoted by $\cI_C$. This coarse image is used as a conditioning input to a Stable Diffusion model \cite{rombach2022} at the RX. 

At \ac{RX}, the caption is reconstructed from the received packet group \(\hat{\mathcal{G}}\) using various \ac{PA} methods. The resulting reconstructed caption, \(\widehat{\cW}\), is then combined with the coarse image and processed by the diffusion model to generate the final image:

\begin{equation}
    \cI_R = S\bigl(\cI_C,\, \hat{\mathcal{G}}\bigr),
\end{equation}
where $S(\cdot,\cdot)$ represents the image synthesis process via Stable Diffusion, with $\cI_C$ serving as the visual guide and $\hat{\mathcal{G}}$ encoding the semantic information from the transmitted caption.

The fidelity of the reconstructed image $\cI_R$ compared to the original image $\cI$ is quantified using the learned perceptual image patch similarity (LPIPS) metric:
\begin{align}
    \sum_{l} \frac{1}{H_l W_l} \sum_{i,j} \bigl\| f_{l}(\cI_R) - f_{l}(\cI) \bigr\|_{2}^{2},
\end{align}
where $f_{l}(\cdot)$ denotes the feature map extracted from the $l$th layer of a pre-trained AlexNet, and $H_l$, $W_l$ are the spatial dimensions of the corresponding feature map. A lower LPIPS value indicates higher perceptual similarity.

Fig.~\ref{fig:plot_word_LPIPS} and Fig.~\ref{fig:plot_token_LPIPS} show the average LPIPS between the images generated from the reconstructed captions and the original images as $p$ increases. 
Here, our proposed method, \ac{SemPA-GBeam} shows LPIPS scores that are very competitive with full search and notably lower than Random \ac{PA}. 

Furthermore, as illustrated in Fig.~\ref{fig:plot_p}, retaining a higher level of semantic content in the reconstructed caption directly corresponds to a decrease in the LPIPS score, thereby yielding images that more closely resemble the original. Such correlation highlights the ability of \ac{SemPA-GBeam} to effectively preserve critical semantic information under packet loss, ultimately enhancing the quality of the generated images.

With word-based tokenization, the caption ``The cat that hides inside a small basket" becomes ``cat that inside a small basket" using \ac{SemPA-GBeam}, preserving key tokens like ``cat" and ``basket" despite packet losses. In contrast, random \ac{PA} produces ``The that hides inside a small," losing critical tokens simultaneously and resulting in images lacking these essential visual elements.

For subword-based tokenization, \ac{SemPA-GBeam} reconstructs ``The cat that sitting near a sneaker" as ``The cat that near sneaker," maintaining its core meaning. Conversely, random \ac{PA} yields ``The cat that sitting near \#er," losing ``sneak\#" from "sneaker" due to packet errors, leading to the absence of this important object. Resulting images highlight these semantic differences clearly, with \ac{SemPA-GBeam} producing images closely resembling originals, while random \ac{PA} generates images with significant visual distortions or missing objects.

\section{Conclusions}
In this paper, we introduced the \ac{SemPA-GBeam} method for text-based semantic communication, which achieves near-optimal performance—comparable to Full Search—in terms of text similarity and LPIPS, while significantly reducing computational complexity. Our approach effectively mitigates semantic errors over unreliable channels. Although our method substantially cuts down the number of text encoding steps, its complexity still grows exponentially with the number of packets \(N\) (i.e., \(2^N\) text encoding steps). 

Future work will focus on developing even more scalable algorithms that further lower this computational burden. In addition, designing more realistic packet-level channel models will be critical for further enhancing the preservation of semantic content in dynamic communication environments.

\bibliographystyle{IEEEtran}
\bibliography{si}

\begin{thebibliography}{10}
\providecommand{\url}[1]{#1}
\csname url@samestyle\endcsname
\providecommand{\newblock}{\relax}
\providecommand{\bibinfo}[2]{#2}
\providecommand{\BIBentrySTDinterwordspacing}{\spaceskip=0pt\relax}
\providecommand{\BIBentryALTinterwordstretchfactor}{4}
\providecommand{\BIBentryALTinterwordspacing}{\spaceskip=\fontdimen2\font plus
\BIBentryALTinterwordstretchfactor\fontdimen3\font minus \fontdimen4\font\relax}
\providecommand{\BIBforeignlanguage}[2]{{%
\expandafter\ifx\csname l@#1\endcsname\relax
\typeout{** WARNING: IEEEtran.bst: No hyphenation pattern has been}%
\typeout{** loaded for the language `#1'. Using the pattern for}%
\typeout{** the default language instead.}%
\else
\language=\csname l@#1\endcsname
\fi
#2}}
\providecommand{\BIBdecl}{\relax}
\BIBdecl

\bibitem{Minrui24}
M.~Xu, H.~Du, D.~Niyato \emph{et~al.}, ``Unleashing the power of edge-cloud generative ai in mobile networks: A survey of {AIGC} services,'' \emph{IEEE Commun. Surv. Tutor.}, vol.~26, no.~2, pp. 1127--1170, 2024.

\bibitem{oh2025HLM}
\BIBentryALTinterwordspacing
S.~Oh, J.~Kim, J.~Park, S.-W. Ko, T.~Q.~S. Quek, and S.-L. Kim, ``Uncertainty-aware hybrid inference with on-device small and remote large language models,'' 2025. [Online]. Available: \url{https://arxiv.org/abs/2412.12687}
\BIBentrySTDinterwordspacing

\bibitem{qiao2025tc}
\BIBentryALTinterwordspacing
L.~Qiao, M.~B. Mashhadi, Z.~Gao, R.~Tafazolli, M.~Bennis, and D.~Niyato, ``Token communications: A unified framework for cross-modal context-aware semantic communications,'' 2025. [Online]. Available: \url{https://arxiv.org/abs/2502.12096}
\BIBentrySTDinterwordspacing

\bibitem{lee2025sempa}
\BIBentryALTinterwordspacing
S.~Lee, J.~Park, J.~Choi, and H.~Park, ``Semantic packet aggregation and repeated transmission for text-to-image generation,'' (to be presented at ICC 2025), 2025. [Online]. Available: \url{https://arxiv.org/abs/2503.23734}
\BIBentrySTDinterwordspacing

\bibitem{MSCOCO14}
T.-Y. Lin, M.~Maire, S.~Belongie \emph{et~al.}, ``{Microsoft coco}: Common objects in context,'' in \emph{Proc. ECCV, Zurich, Switzerland, September}, 2014.

\bibitem{Qin22}
L.~Yan, Z.~Qin, R.~Zhang, Y.~Li, and G.~Y. Li, ``Resource allocation for text semantic communications,'' \emph{IEEE Wireless Communications Letters}, vol.~11, no.~7, pp. 1394--1398, 2022.

\bibitem{Xidong23}
X.~Mu, Y.~Liu, L.~Guo, and N.~Al-Dhahir, ``Heterogeneous semantic and bit communications: A semi-{NOMA} scheme,'' \emph{IEEE Journal on Selected Areas in Communications}, vol.~41, no.~1, pp. 155--169, 2023.

\bibitem{Jinsong24}
J.~Hu, L.~Ye, Y.~Chen, X.~Zhang, J.~Wang, and Z.~Chen, ``Covert communications for text semantic with finite blocklength,'' \emph{IEEE Wireless Communications Letters}, pp. 1--1, 2024.

\bibitem{Nam2023LanguageOrientedCW}
H.~Nam, J.~Park, J.~Choi, M.~Bennis, and S.-L. Kim, ``Language-oriented communication with semantic coding and knowledge distillation for text-to-image generation,'' in \emph{ICASSP 2024}, 2024, pp. 13\,506--13\,510.

\bibitem{Guo22}
S.~Guo, Y.~Wang, S.~Li, and N.~Saeed, ``Semantic importance-aware communications using pre-trained language models,'' \emph{IEEE Communications Letters}, vol.~27, no.~9, pp. 2328--2332, 2023.

\bibitem{Guo24}
\BIBentryALTinterwordspacing
S.~Guo, Y.~Wang, J.~Ye, A.~Zhang, and K.~Xu, ``Semantic importance-aware communications with semantic correction using large language models,'' 2024. [Online]. Available: \url{https://arxiv.org/abs/2405.16011}
\BIBentrySTDinterwordspacing

\bibitem{cicchetti24}
G.~Cicchetti, E.~Grassucci, J.~Park, J.~Choi, S.~Barbarossa, and D.~Comminiello, ``Language-oriented semantic latent representation for image transmission,'' 2024.

\bibitem{Sundararajan11}
J.~K. Sundararajan, D.~Shah, M.~Médard, S.~Jakubczak, M.~Mitzenmacher, and J.~Barros, ``Network coding meets {TCP}: Theory and implementation,'' \emph{Proceedings of the IEEE}, vol.~99, no.~3, pp. 490--512, 2011.

\bibitem{CLIP}
A.~Radford, J.~W. Kim, C.~Hallacy \emph{et~al.}, ``Learning transferable visual models from natural language supervision,'' in \emph{International conference on machine learning}.\hskip 1em plus 0.5em minus 0.4em\relax PMLR, 2021, pp. 8748--8763.

\bibitem{katoch21}
S.~Katoch, S.~S. Chauhan, and V.~Kumar, ``A review on genetic algorithm: past, present, and future,'' \emph{Multimedia tools and applications}, vol.~80, pp. 8091--8126, 2021.

\bibitem{snell24}
\BIBentryALTinterwordspacing
C.~Snell, J.~Lee, K.~Xu, and A.~Kumar, ``Scaling llm test-time compute optimally can be more effective than scaling model parameters,'' 2024. [Online]. Available: \url{https://arxiv.org/abs/2408.03314}
\BIBentrySTDinterwordspacing

\bibitem{rombach2022}
R.~Rombach, A.~Blattmann, D.~Lorenz, P.~Esser, and B.~Ommer, ``High-resolution image synthesis with latent diffusion models,'' in \emph{Proceedings of the IEEE/CVF conference on computer vision and pattern recognition}, 2022, pp. 10\,684--10\,695.

\end{thebibliography}
\end{document}